\documentclass[runningheads]{llncs}
\usepackage[T1]{fontenc}
\usepackage{newtxtext}       %
\usepackage[varvw]{newtxmath}      
\usepackage{microtype,booktabs,tabularx,array,enumitem,needspace}
\usepackage{graphicx}
\usepackage{hyperref}
\newcommand{\E}{\mathbb E}
\newcommand{\Prob}{\mathbb P}
\newcommand{\OPT}{\operatorname{OPT}}
\newcommand{\Bin}{\operatorname{Bin}}
\newcommand{\ind}{\mathbf 1}

\newcommand{\K}{\mathsf{K}}
\newcommand{\eps}{\varepsilon}
\newcommand{\vhat}{\widehat v}

\numberwithin{equation}{section}
\usepackage{etoolbox}
\AtEndEnvironment{proof}{\qed}
\begin{document}
\title{Multiple-Choice Secretary Problem with Predictions}
\author{Jing Yuan\inst{1}\and  Qi Cai\inst{1} \and Xin Gao\inst{1} \and
Shaojie Tang\inst{2}}
\institute{University of North Texas \and
University at Buffalo}
\maketitle             
\begin{abstract}
We study the multiple-choice secretary problem, in which candidates
arrive in uniformly random order and the goal is to select at most
$k$ candidates to maximize their total value. We consider a setting
where predictions of individual candidate values are available
before any candidate arrives. Fujii and Yoshida's Learned Kleinberg algorithm achieves competitive loss $O(\min\{\log k/\sqrt{k},\eps\})$, where $k$ is the selection capacity and $\eps$ is the maximum multiplicative prediction error. We remove the logarithmic factor through a sharper analysis of the same switching algorithm with a retuned threshold. The resulting competitive ratio is at least $\max\{0,1-\min\{32/\sqrt{k},4\eps\}\}$. In particular, exact  predictions ($\eps=0$) achieve the offline
optimum on every arrival order.  
\end{abstract}

\section{Introduction}\label{sec:intro}

Online algorithms make decisions under uncertainty about future
inputs. Advances in machine learning allow predictions to guide
these decisions, raising the question of how to exploit accurate
predictions while remaining robust to errors. The multiple-choice secretary problem models online selection
under a capacity constraint. Candidates arrive in uniformly random
order, and each candidate's value is revealed upon arrival.
The algorithm must immediately accept or reject each candidate,
selecting at most $k$ to maximize their total value.
The challenge is to balance accepting valuable candidates now
against reserving capacity for better candidates later.
Without predictions, Kleinberg~\cite{kleinberg2005} achieves
competitive ratio $1-O(k^{-1/2})$ against the sum of the $k$
largest values. With predictions, we additionally seek
\emph{perfect consistency}: recovering the offline optimum
when predictions are exact, while retaining a strong guarantee
under arbitrary errors. These objectives follow the
consistency--robustness framework~\cite{mitzenmacher2022}.

Fujii and Yoshida~\cite{fujii2024} introduced the candidate-value
prediction model studied here, in which predictions of all
candidates' values are available before arrivals. A motivating
application is hiring: machine learning models can estimate
candidates' abilities from their r\'esum\'es or publicly available
information before interviews. These estimates may be inaccurate
because the information is incomplete, misleading, or deliberately
manipulated. Interviews reveal the candidates' true values, and
the decision maker must immediately accept or reject each candidate,
as an offer may no longer be possible later. The model assumes
that candidates arrive in uniformly random order, as in the classical
secretary problem. Their Learned Kleinberg algorithm follows the predicted top-$k$ set until it observes a sufficiently large error, then switches to a classical secretary algorithm. They prove competitive loss $O(\min\{\log k/\sqrt{k},\eps\})$, where $\eps$ is the maximum multiplicative prediction error, and ask whether the logarithmic factor can be removed.

\paragraph{Our result.}
We resolve this question by proving loss $O(\min\{k^{-1/2},\eps\})$ for the same switching algorithm with a different monitoring threshold. This algorithm does not need to know $\eps$. The explicit guarantee is
\[
\E[v(A)]\ge
\max\left\{0,1-\min\left\{\frac{32}{\sqrt{k}},4\eps\right\}\right\}\OPT.
\]
Our algorithm matches the classical $\Omega(k^{-1/2})$ lower bound
on worst-case loss~\cite{kleinberg2005,qiao2026} up to a constant
factor, while recovering the offline optimum on every arrival
order when predictions are exact. Table~\ref{tab:comparison} places the result alongside the closest multiple-choice guarantees.

\begin{table}[ht]
\centering\small
\setlength{\tabcolsep}{4pt}
\begin{tabularx}{\textwidth}{@{}p{0.3\textwidth}p{0.2\textwidth}X@{}}
\toprule
Work & Prediction & Competitive guarantee\\
\midrule
Kleinberg~\cite{kleinberg2005} & None & $1-O(k^{-1/2})$\\[3pt]
Qiao--Zhang~\cite{qiao2026} & None & $1-O(k^{-1/2})$ with $O(\log k)$ memory\\[3pt]
Fujii--Yoshida~\cite{fujii2024} & Candidate values & $1-O(\min\{\log k/\sqrt{k},\eps\})$\\[3pt]
This paper & Candidate values & $1-O(\min\{k^{-1/2},\eps\})$\\
\bottomrule
\end{tabularx}
\caption{Competitive ratios for random-order multiple-choice selection.}\label{tab:comparison}
\end{table}

\section{Problem and algorithm}\label{sec:model}

\subsection{Model and prediction error}

Let $N=\{1,\ldots,n\}$ and $1\le k\le n$. Candidate $i$ has a nonnegative value $v_i$ and a nonnegative prediction $\vhat_i$. Both vectors are fixed before a uniformly random permutation of $N$ is drawn. The algorithm initially knows $n$, $k$, the labels, and all predictions. It observes $v_i$ when $i$ arrives and must accept or reject $i$ immediately. Acceptances are irrevocable, and at most $k$ are allowed.

For $D\subseteq N$, write $v(D)=\sum_{i\in D}v_i$ and $\vhat(D)=\sum_{i\in D}\vhat_i$. Fix a label-based rule for breaking ties. Let $P$ be the predicted top-$k$ set and $O$ the true top-$k$ set, so $\OPT=v(O)$. The prediction error is $\eps=\max_i e_i$, where
\[
e_i=
\begin{cases}
|\vhat_i-v_i|/v_i,&v_i>0,\\
0,&v_i=\vhat_i=0,\\
+\infty,&v_i=0<\vhat_i.
\end{cases}
\]
For finite $\eps$,
\begin{equation}\label{eq:errorband}
(1-\eps)v_i\le\vhat_i\le(1+\eps)v_i.
\end{equation}
A competitive guarantee $g(k,\eps)$ means $\E[v(A)]\ge g(k,\eps)\OPT$ on every fixed input, with expectation over arrival order and internal randomness. Following the consistency--robustness framework for algorithms
with predictions~\cite{mitzenmacher2022,fujii2024},
perfect consistency means achieving $\OPT$ when $\eps=0$,
while robustness requires a competitive guarantee that holds
for every prediction vector, regardless of its accuracy. 

\subsection{Algorithm design}
\label{sec:algorithm}
For a fixed candidate set $Z$, let $\OPT_q(Z)$ be the sum of its largest $\min\{q,|Z|\}$ values, with $\OPT_0(Z)=0$. For the classical multiple-choice secretary problem without
predictions, Kleinberg~\cite{kleinberg2005}
provides a randomized algorithm $\K_q$ that selects at most
$q$ candidates from a uniformly ordered stream $Z$ and satisfies
\begin{equation}\label{eq:kleinberg}
\E[v(\K_q(Z))]\ge c_q\OPT_q(Z),
\qquad c_q=\max\{0,1-5/\sqrt q\},\quad q\ge1.
\end{equation}
For zero capacity or an empty stream, it selects nothing;
when $q\ge|Z|$, it accepts all remaining candidates. We use $\K_q$ as a black-box fallback in the algorithm below.

Specifically, we adopt Fujii and Yoshida's Learned Kleinberg
algorithm~\cite{fujii2024} with a retuned threshold $\theta$,
as shown in Algorithm~\ref{alg:learned}.
It accepts candidates in the predicted top-$k$ set $P$ until
the first prediction error exceeds $\theta$ or capacity is exhausted.
If capacity remains, it accepts the candidate that triggers the
switch and runs Kleinberg's algorithm $\K_q$ ~\cite{kleinberg2005}
on subsequent arrivals with the remaining slots.

\newcounter{algorithm}
\refstepcounter{algorithm}\label{alg:learned}
\begin{center}
\begin{minipage}{0.96\textwidth}
\hrule\vspace{6pt}
\textbf{Algorithm \thealgorithm. Learned Kleinberg with threshold $\theta$ ~\cite{fujii2024}}\\
\textit{Input:} predictions $\vhat$, capacity $k$, and $0\le\theta<1$.
\begin{enumerate}
\item Compute the predicted top-$k$ set $P$ and initialize $A=\varnothing$.
\item On arrival of $i$, stop if $|A|=k$.
\item If $|\vhat_i-v_i|>\theta v_i$, add $i$ to $A$, set $q=k-|A|$, and run a fresh $\K_q$ on all subsequent arrivals. Return $A$ together with the fallback $\K_q$'s selections.
\item Otherwise, add $i$ to $A$ if $i\in P$, and continue.
\end{enumerate}
\vspace{2pt}\hrule
\end{minipage}
\end{center}

\subsection{Main guarantees}

\begin{theorem}\label{thm:main}
With threshold
\begin{equation}\label{eq:theta}
\theta=\min\left\{\frac12,\frac8{\sqrt{k}}\right\},
\end{equation}
Algorithm~\ref{alg:learned} satisfies
\begin{equation}\label{eq:main}
\E[v(A)]\ge
\max\left\{0,1-\min\left\{\frac{32}{\sqrt{k}},4\eps\right\}\right\}\OPT.
\end{equation}
When $\eps=0$, it obtains $\OPT$ on every arrival order.
\end{theorem}

The result follows from the next threshold-dependent bound and the accuracy guarantee of selecting $P$ when no error triggers a switch.

\begin{theorem}\label{thm:robust}
For every $0\le\theta<1$, Algorithm~\ref{alg:learned} satisfies
\begin{equation}\label{eq:robust}
\E[v(A)]\ge\left(1-2\theta-\frac{16}{\sqrt{k}}\right)\OPT.
\end{equation}
\end{theorem}

The remainder of the paper develops the analysis needed to prove
Theorems~\ref{thm:robust} and~\ref{thm:main}.

\section{Proof overview and stopping-time setup}\label{sec:setup}

Fix a threshold $\theta$ and let
\begin{equation}\label{eq:M}
M=\{i:|\vhat_i-v_i|>\theta v_i\},\qquad m=|M|.
\end{equation}
The set $M$ is fixed before the arrival permutation. If $m=0$, the algorithm selects $P$, and
\begin{equation}\label{eq:noerror}
v(P)\ge\frac{\vhat(P)}{1+\theta}
\ge\frac{\vhat(O)}{1+\theta}
\ge\frac{1-\theta}{1+\theta}\OPT
\ge(1-2\theta)\OPT.
\end{equation}
We therefore focus on $m\ge1$.

\paragraph{The fixed benchmark.}
Let
\begin{equation}\label{eq:comparison}
B=(P\setminus M)\cup(O\cap M),\qquad
\alpha=\frac{1-\theta}{1+\theta}.
\end{equation}
The set $B$ combines the accurately predicted part of $P$ with the erroneous part of the optimum. Its value is close to $\OPT$, and all its members arriving before the first error are accepted.  The proof has three steps:
\begin{enumerate}
\item Compare $v(B)$ with $\OPT$.
\item Condition on the switch and bound prefix value plus expected fallback value by a capacity-dependent fraction of $v(B)$.
\item Average that fraction over the remaining capacity.
\end{enumerate}

\paragraph{Representing the random order.}
Assign independent times $U_i\sim\operatorname{Unif}[0,1]$ and process candidates in increasing time order. This standard representation~\cite{bruss1984,kleinberg2005} also underlies Fujii and Yoshida's analysis~\cite{fujii2024}. Define
\begin{equation}\label{eq:trigger}
T=\min_{i\in M}U_i,\qquad J=\arg\min_{i\in M}U_i,
\end{equation}
and
\begin{equation}\label{eq:prefix}
S=P\cap\{i:U_i<T\},\quad s=|S|,\quad r=k-s,\quad
Z=\{i:U_i>T\}.
\end{equation}
Here $S$ is the set of candidates accepted before the trigger candidate $J$
arrives, $r$ is the number of unfilled selection slots immediately
before  $J$'s arrival, and $Z$ is the suffix. For $r>0$, the algorithm accepts $J$ and gives $q=r-1$ slots to the fallback. For $r=0$, it has already accepted all of $P$.

Let $\mathcal H$ record the ordered prefix through the trigger,
the trigger time $T$ and identity $J$, and the remaining set $Z$.
Conditional on $\mathcal H$, the arrival times of candidates in
$Z$ are independent and uniform on $(T,1)$, so their relative
order is uniform. Thus, for $r\ge2$, the fallback guarantee
in~\eqref{eq:kleinberg} applies with capacity $q=r-1$;
for $r\le1$, no capacity remains for the fallback.

Since $T$ is the earliest arrival time among the $m$ erroneous
candidates, whose arrival times are independent and uniform
on $[0,1]$, $\Prob(T>t)=(1-t)^m, 0\le t\le1$. Consequently, the remaining-time fraction $p:=1-T$ satisfies
$\Prob(p\le u)=u^m, 0\le u\le1$, and has density $mu^{m-1}$ on $(0,1)$.  

Following the error-count decomposition in Fujii and Yoshida~\cite{fujii2024}, we distinguish two regimes, with the split now at $\sqrt k$. When $m>\sqrt k$, the first error arrives early enough in expectation that little value or capacity has been lost. When $1\le m\le\sqrt k$, the switch may occur late, and the value collected before switching becomes essential. 

\section{Few erroneous candidates: $1\le m\le\sqrt k$}\label{sec:few}

We first consider the few-error case, $1\le m\le\sqrt{k}$.
We show that the benchmark set $B$ retains most of the optimal
value, then bound how much of this value the algorithm recovers
across its two phases.

Recall that every $i\notin M$ satisfies
\begin{equation}\label{eq:goodband}
(1-\theta)v_i\le\vhat_i\le(1+\theta)v_i.
\end{equation}
The next lemma adapts the value comparison in the proof of
Lemma~6 of Fujii and Yoshida~\cite{fujii2024}
to bound $v(B)$ in terms of the number $m$ of erroneous candidates.

\begin{lemma}\label{lem:Bvalue}
The set in~\eqref{eq:comparison} satisfies
\begin{equation}\label{eq:Bvalue}
v(B)\ge\alpha\left(1-\frac m k\right)\OPT.
\end{equation}
\end{lemma}
\begin{proof}
Let $a=|P\setminus M|$ and $d=|O\setminus M|$. Removing $M$ from a predicted top-$k$ set leaves the top $a$ predictions among $N\setminus M$. To see this, any candidate in $P\setminus M$ has predicted rank ahead of every candidate outside $P$, under the fixed tie rule. 

If $d>0$, the top-$a$ predicted sum in $N\setminus M$ is at least $\min\{1,a/d\}\vhat(O\setminus M)$. When $a\ge d$, take the best $d$ predictions and add nonnegative terms. When $a<d$, the largest $a$ predictions within $O\setminus M$ have average at least its overall average. Since $a\ge k-m$, $d\le k$, and $m\le k$, this proves $\vhat(P\setminus M)\ge(1-m/k)\vhat(O\setminus M)$. Applying~\eqref{eq:goodband} to both sets yields $v(P\setminus M)\ge\alpha(1-m/k)v(O\setminus M)$. This inequality also holds when $d=0$, since its right-hand
side is zero. Adding $v(O\cap M)$ and using
$0\le\alpha(1-m/k)\le1$ gives
$v(B)\ge\alpha(1-m/k)v(O)=\alpha(1-m/k)\OPT$,
as claimed.
\end{proof}

To turn this comparison into a performance bound, we must account for the fact that a late switch may leave only a few slots. The next result makes the value already collected compensate for that loss of capacity.

Define
\begin{equation}\label{eq:g}
g_m(r)=
\begin{cases}
0,&r\le1,\\
\displaystyle c_{r-1}\frac{r-1}{r+m},&r\ge2,
\end{cases}
\end{equation}
where $c_q$ is the fallback ratio in~\eqref{eq:kleinberg}. The factor $c_{r-1}$ accounts for the classical fallback, and $(r-1)/(r+m)$ accounts for the benchmark's possible excess size.

\begin{lemma}\label{lem:conditionalB}
For the conditioning information $\mathcal H$ defined in Section~\ref{sec:setup},
\begin{equation}\label{eq:conditionalB}
\E[v(A)\mid\mathcal H]\ge g_m(r)v(B).
\end{equation}
Consequently, $\E[v(A)]\ge\E[g_m(r)]v(B)$.
\end{lemma}
\begin{proof}
When $r\le1$, the right side is zero. Suppose $r\ge2$, and put $q=r-1$. Recall that $S=P\cap\{i:U_i<T\}$ is the set of candidates
accepted before the first erroneous candidate arrives at time $T$.  All members of $B$ arriving before $T$ belong to $P\setminus M$, so they are in $S$. The trigger is also accepted. Thus
\begin{equation}\label{eq:C}
C=B\setminus(S\cup\{J\})\subseteq Z.
\end{equation}
Write $b=|P\cap M|$ and $c=|O\cap M|$. Since $|P\setminus M|=k-b$,
\[
|C|=k-b-s+c-\ind_{\{J\in O\}}
=q+1+c-b-\ind_{\{J\in O\}}\le q+m+1.
\]
The largest $q$ values of $C$, or all of $C$ if it is smaller, therefore give
\begin{equation}\label{eq:completion}
\OPT_q(Z)\ge\frac{q}{q+m+1}v(C).
\end{equation}
The fallback contributes at least $g_m(r)v(C)$ conditionally on $\mathcal H$. The already accepted candidates contribute
$v(S)+v_J\ge v(B\setminus C)$. Since $0\le g_m(r)\le1$,
\[
\E[v(A)\mid\mathcal H]
\ge v(B\setminus C)+g_m(r)v(C)
\ge g_m(r)v(B).
\]
Finally, $B$ is fixed before the arrival times are sampled, so $v(B)$ can be taken outside the expectation.
\end{proof}
\subsection{Lower bounding $\E[g_m(r)]$}\label{sec:capacity}

Lemma~\ref{lem:conditionalB} reduces the few-error analysis to bounding $\E[g_m(r)]$. We first compare $r$ with a binomial mixture, then average the two losses in~\eqref{eq:g}.

We use the same conditional binomial description of the predicted candidates as in Fujii and Yoshida~\cite[Lemma~3]{fujii2024}. Let $b=|P\cap M|$. Conditional on $p=1-T$, all $b$ erroneous members of $P$ are still unprocessed just before $T$. Each of its other $k-b$ members remains independently with probability $p$. Hence
\begin{equation}\label{eq:rbinomial}
r\mid p\ \stackrel{d}{=}\ b+\Bin(k-b,p).
\end{equation}
To obtain a simpler lower bound on $r$, condition on $p$ and let
$B_1,\ldots,B_k$ be independent Bernoulli random variables, each equal
to $1$ with probability $p$. We can represent
$r=b+\sum_{i=b+1}^{k}B_i$ and define $X=\sum_{i=1}^{k}B_i$.
Since $\sum_{i=1}^{b}B_i\le b$, this construction ensures
\begin{equation}\label{eq:coupling}
r\ge X,\qquad X\mid p\sim\Bin(k,p).
\end{equation}
Thus, conditional on $p$, this capacity lower bound $X$ has a binomial distribution with $k$
trials and success probability $p$.

Fujii and Yoshida~\cite{fujii2024}
bound a reciprocal-capacity expectation by comparing it with the case
$m=1$. We use the same reference case, make the comparison explicit
through a coupling, and apply it directly to the clipped square-root
loss.

Let $V$ be uniform on $[0,1]$. When $m=1$, the remaining-time fraction
$p=1-T$ is uniform, so we define the reference variable by
$X_0\mid V\sim\Bin(k,V)$. Averaging over $V$ gives
\[
\Prob(X_0=x)
=\int_0^1 \binom{k}{x}v^x(1-v)^{k-x}\,dv
=\frac{1}{k+1},
\qquad x=0,\ldots,k.
\]
Thus, although $X_0$ is binomial conditional on $V$, its unconditional
distribution is uniform on $\{0,\ldots,k\}$.

For general $m\ge1$, retain $p=1-T$ and define $V:=p^m$.
Since $\Prob(p\le u)=u^m$, the variable $V$ is uniform
on $[0,1]$, and $p=V^{1/m}\ge V$.
To compare the two conditional binomial distributions $X$ and $X_0$,
take auxiliary independent uniform random variables
$W_1,\ldots,W_k$, independent of $p$, and set
$X=\sum_{i=1}^k\ind_{\{W_i\le V^{1/m}\}},
X_0=\sum_{i=1}^k\ind_{\{W_i\le V\}}$.

These variables have the required conditional binomial distributions.
Since $V^{1/m}\ge V$, every success counted by $X_0$ is also counted by $X$,
and hence $X\ge X_0$. Consequently, for every nonincreasing function
$f:\{0,\ldots,k\}\to\mathbb{R}$, we have $f(X)\le f(X_0)$ and therefore
\begin{equation}\label{eq:monotone}
\E[f(X)]
\le \E[f(X_0)]
=\frac{1}{k+1}\sum_{x=0}^k f(x).
\end{equation}

We now derive a lower bound on $\E[g_m(r)]$.

\begin{lemma}\label{lem:averaging}
For $1\le m\le\sqrt k$,
\begin{equation}\label{eq:Eg}
\E[g_m(r)]\ge1-\frac{15}{\sqrt k}.
\end{equation}
\end{lemma}
\begin{proof}
There are two losses in~\eqref{eq:g}: the fraction of the remaining benchmark that fits, and the classical fallback's approximation loss. Define
\begin{equation}\label{eq:h}
h(x)=
\begin{cases}
1,&x=0,1,\\
\min\{1,5/\sqrt{x-1}\},&x\ge2.
\end{cases}
\end{equation}
Using $r\ge X$ from~\eqref{eq:coupling},
\begin{equation}\label{eq:loss}
1-g_m(r)\le\frac{m+1}{X+m}+h(X).
\end{equation}
Indeed, for $X\ge2$, write $g_m(r)=(1-u)(1-w)$ with
$u=(m+1)/(r+m)$ and $w=\min\{1,5/\sqrt{r-1}\}$.
Then $1-g_m(r)\le u+w$, and $r\ge X$ gives~\eqref{eq:loss}. For $X\le1$, the right side is at least one.

The function $h$ is nonincreasing. Applying~\eqref{eq:monotone} directly to this square-root penalty yields
\begin{equation}\label{eq:sqrtloss}
\E[h(X)]\le
\frac{2+5\sum_{j=1}^{k-1}j^{-1/2}}{k+1}
\le\frac{11}{\sqrt k}.
\end{equation} The last inequality uses
$\sum_{j=1}^{k-1}j^{-1/2}\le2\sqrt{k-1}$,
$2/(k+1)\le1/\sqrt{k}$, and
$\sqrt{k-1}/(k+1)\le1/\sqrt{k}$,
with an empty sum when $k=1$.
We bound the capacity-excess term separately for $m=1$ and
$2\le m\le\sqrt{k}$. When $m=1$, $X$ is uniform on
$\{0,\ldots,k\}$, so
\[
\E\left[\frac{2}{X+1}\right]
=\frac{2}{k+1}\sum_{x=0}^{k}\frac1{x+1}
=\frac{2H_{k+1}}{k+1}
\le\frac4{\sqrt{k}},
\]
where $H_\ell=\sum_{j=1}^{\ell}1/j$ and we used
$H_\ell\le2\sqrt{\ell}$.

For $m\ge2$, we use the classical binomial reciprocal
identity~\cite{chao1972}, following Fujii and
Yoshida~\cite[Lemma~4]{fujii2024}.
Since $X\mid p\sim\Bin(k,p)$,
\[
\E\left[\frac1{X+1}\,\middle|\,p\right]
=\frac{1-(1-p)^{k+1}}{(k+1)p}
\le\frac1{(k+1)p}.
\]
Moreover, $p=1-T$ has density $mp^{m-1}$, and hence
$\E\left[\frac1p\right]
=\int_0^1 mp^{m-2}\,dp
=\frac{m}{m-1}$.

Using $X+m\ge X+1$ and averaging over $p$, we obtain
\begin{equation}\label{eq:capacityloss}
\begin{aligned}
\E\left[\frac{m+1}{X+m}\right]
&\le (m+1)\E\left[\frac1{X+1}\right]\le \frac{m+1}{k+1}\E\left[\frac1p\right]
=\frac{m(m+1)}{(k+1)(m-1)}\\
&\le\frac{3m}{k+1}
\le\frac3{\sqrt{k}}
\le\frac4{\sqrt{k}}.
\end{aligned}
\end{equation}
Here $(m+1)/(m-1)\le3$ for $m\ge2$, and
$m\le\sqrt{k}$. Combining~\eqref{eq:loss}--\eqref{eq:capacityloss} proves the lemma.
\end{proof}

\subsection{Combining the few-error bounds}

\begin{corollary}\label{cor:few}
If $1\le m\le\sqrt{k}$, then
$\E[v(A)]
\ge\left(1-2\theta-\frac{16}{\sqrt{k}}\right)\OPT$.
\end{corollary}
\begin{proof}
If $1-2\theta-16/\sqrt{k}\le0$, the claim follows from
nonnegativity. Otherwise, $1-15/\sqrt{k}>0$.
Lemmas~\ref{lem:conditionalB} and~\ref{lem:averaging} give
$\E[v(A)]
\ge\E[g_m(r)]\,v(B)
\ge\left(1-\frac{15}{\sqrt{k}}\right)v(B)$.

Applying Lemma~\ref{lem:Bvalue}, we obtain
\begin{align}
\E[v(A)]
&\ge
\left(1-\frac{15}{\sqrt{k}}\right)
\frac{1-\theta}{1+\theta}
\left(1-\frac{m}{k}\right)\OPT \notag\\
&\ge
\left(1-\frac{15}{\sqrt{k}}
-\frac{2\theta}{1+\theta}-\frac{m}{k}\right)\OPT \notag\\
&\ge
\left(1-2\theta-\frac{16}{\sqrt{k}}\right)\OPT.
\label{eq:fewfinal}
\end{align}
The second inequality uses
$(1-a)(1-b)(1-c)\ge1-a-b-c$ for $a,b,c\in[0,1]$.
The last inequality follows from
$2\theta/(1+\theta)\le2\theta$ and $m/k\le1/\sqrt{k}$.
\end{proof}

\section{Many erroneous candidates: $m>\sqrt k$}\label{sec:many}

We now consider the many-error case, $m>\sqrt{k}$, in which the
algorithm tends to switch early. Following the early-switch
analysis of Fujii and Yoshida~\cite{fujii2024},
we control the losses due to the switch time and the capacity
used before switching. Our argument bounds these losses directly
in expectation, replacing their high-probability bounds.

\begin{lemma}\label{lem:many}
If $m>\sqrt k$, then
\begin{equation}\label{eq:many}
\E[v(A)]\ge\left(1-\frac8{\sqrt k}\right)\OPT.
\end{equation}
\end{lemma}
\begin{proof}
Recall that $Z=\{i:U_i>T\}$ is the set of candidates arriving
strictly after the trigger at time $T$. Let $W=v(O\cap Z)$. When $r\ge2$, the fallback has
$q=k-s-1$ slots. Since $|O\cap Z|\le k$, we have
$\OPT_q(Z)\ge(q/k)W$. Applying~\eqref{eq:kleinberg}
conditionally on $\mathcal H$ gives
\begin{align}
\E[v(A)\mid\mathcal H]
&\ge\left(\frac qk-\frac{5\sqrt q}{k}\right)W\notag=W-\left(\frac{s+1}{k}+\frac{5\sqrt q}{k}\right)W\notag\\
&\ge W-\left(\frac{s+1}{k}+\frac5{\sqrt k}\right)\OPT,
\label{eq:manyconditional}
\end{align}
using $W\le\OPT$ and $q\le k$.
For $r\le1$, we have $s+1\ge k$, so the final right-hand
side is nonpositive and the same bound follows from nonnegativity.

A candidate in $O\setminus M$ is absent from $Z$ with probability $1/(m+1)$; a candidate in $O\cap M$ is absent precisely when it is the trigger, with probability $1/m$. Similarly, each member of $P\setminus M$ precedes the trigger with probability $1/(m+1)$. Thus
\begin{align}
\E[W]&=\frac{m}{m+1}v(O\setminus M)
+\left(1-\frac1m\right)v(O\cap M)
\ge\left(1-\frac1m\right)\OPT,\label{eq:Wsuffix}\\
\E[s]&=\frac{|P\setminus M|}{m+1}\le\frac{k}{m+1}.\label{eq:expectedS}
\end{align}
Taking expectations in~\eqref{eq:manyconditional} gives
\[
\frac{\E[v(A)]}{\OPT}
\ge1-\frac1m-\frac1{m+1}-\frac1k-\frac5{\sqrt k}
\ge1-\frac8{\sqrt k},
\]
using $m>\sqrt k$.
\end{proof}

\section{Completing the main proof}\label{sec:mainproof}

We combine the few-error analysis with the many-error analysis to prove
Theorem~\ref{thm:robust}. We then choose the threshold and apply
the no-switch accuracy bound when $\eps\le\theta$ and the
robustness bound when $\eps>\theta$ to establish
Theorem~\ref{thm:main}.

\subsection{Proof of Theorem~\ref{thm:robust}}

\begin{proof}[Proof of Theorem~\ref{thm:robust}]
If $m=0$, equation~\eqref{eq:noerror} gives the stronger
bound $\E[v(A)]\ge(1-2\theta)\OPT$.
If $1\le m\le\sqrt{k}$, the result follows from
Corollary~\ref{cor:few}.
If $m>\sqrt{k}$, Lemma~\ref{lem:many} gives
\[
\E[v(A)]
\ge\left(1-\frac8{\sqrt{k}}\right)\OPT
\ge\left(1-2\theta-\frac{16}{\sqrt{k}}\right)\OPT.
\]
These cases exhaust all possible values of $m$.
\end{proof}

\subsection{Proof of Theorem~\ref{thm:main}}

When all prediction errors are at most $\theta$, the algorithm never leaves the predicted shortlist. The following bound is the consistency argument from~\cite{fujii2024}.

\begin{lemma}~\cite[Lemma~1]{fujii2024}\label{lem:consistent}
If $\eps\le\theta<1$, then $A=P$ and
\begin{equation}\label{eq:consistency}
v(A)\ge\frac{1-\eps}{1+\eps}\OPT\ge(1-2\eps)\OPT.
\end{equation}
\end{lemma}
Now we are ready to prove Theorem~\ref{thm:main}.
\begin{proof}[Proof of Theorem~\ref{thm:main}]
Recall that $\theta=\min\{1/2,8/\sqrt{k}\}$.
We first prove the bound without clipping it at zero.

Suppose $k\ge256$, so $\theta=8/\sqrt{k}$.
Theorem~\ref{thm:robust} gives, for arbitrary predictions,
\[
\E[v(A)]
\ge\left(1-2\theta-\frac{16}{\sqrt{k}}\right)\OPT
=\left(1-\frac{32}{\sqrt{k}}\right)\OPT.
\]
If $\eps>\theta$, then $4\eps>32/\sqrt{k}$, so this is
exactly the desired bound
\[
\E[v(A)]
\ge\left(1-\min\left\{\frac{32}{\sqrt{k}},4\eps\right\}\right)\OPT.
\]
If $\eps\le\theta$, no error exceeds the switching threshold,
and Lemma~\ref{lem:consistent} also gives
$\E[v(A)]\ge(1-2\eps)\OPT$.
Combining the two guarantees yields
\[
\E[v(A)]
\ge\max\left\{1-\frac{32}{\sqrt{k}},1-2\eps\right\}\OPT
\ge\left(1-\min\left\{\frac{32}{\sqrt{k}},4\eps\right\}\right)\OPT.
\]

Now suppose $k<256$, so $\theta=1/2$ and $32/\sqrt{k}>2$.
If $\eps\le1/2$, then
$\min\{32/\sqrt{k},4\eps\}=4\eps$.
Lemma~\ref{lem:consistent} therefore gives
\[
\E[v(A)]\ge(1-2\eps)\OPT\ge(1-4\eps)\OPT.
\]
If $\eps>1/2$, both $32/\sqrt{k}$ and $4\eps$ exceed two,
so the desired bound before clipping is negative and follows
from $v(A)\ge0$.

Finally, nonnegativity allows us to replace the bound by its
maximum with zero, proving~\eqref{eq:main}, including when
$\eps=\infty$.
With exact predictions, no switch occurs and $P$ is a true
top-$k$ set, so
$v(A)=\OPT$ for every arrival order.
\end{proof}

\section{Extension: Measuring prediction error at the selection boundary}
\label{app:relevant}

The maximum prediction error $\eps$ can be large even when the predicted
top-$k$ set $P$ is optimal. For example, if a candidate in $P$ turns out
to be worth more than predicted, this strengthens the case for selecting
that candidate. Yet the original monitoring rule can switch on precisely
such an error. We therefore ask whether the algorithm can retain its
guarantee while ignoring errors of this kind.

Fujii and Yoshida~\cite{fujii2024}
compare candidates outside the predicted top-$k$ set $P$ with its
predicted cutoff $\tau=\min_{i\in P}\vhat_i$, while measuring error
inside $P$ against each candidate's own prediction. We use the same
cutoff on both sides: a shortlisted candidate is a concern only if
its true value falls below $\tau$, and an excluded candidate is a
concern only if its true value exceeds $\tau$.

\paragraph{Cutoff error and switching rule.}
Define the cutoff error by
\begin{equation}\label{eq:boundaryerror}
\eta=\inf\left\{\delta\ge0:
\begin{array}{ll}
\tau\le(1+\delta)v_i & \text{for every }i\in P,\\
\tau\ge(1-\delta)v_i & \text{for every }i\notin P
\end{array}\right\},
\end{equation}
with $\inf\varnothing=+\infty$. In particular, $\eta=0$ precisely
when every shortlisted candidate has value at least $\tau$ and
every excluded candidate has value at most $\tau$. In that case,
$P$ is a true top-$k$ set. \textbf{The cutoff error satisfies $\eta\le\eps$}: for $i\in P$ we have
$\tau\le\vhat_i$, while for $i\notin P$ we have $\tau\ge\vhat_i$.
It is also no larger than Fujii and Yoshida's boundary-sensitive
error whenever their relative errors are defined. 

Modify only the trigger in Algorithm~\ref{alg:learned}. On the
arrival of candidate $i$, switch precisely when
\begin{equation}\label{eq:boundarytrigger}
\bigl(i\in P \text{ and } \tau>(1+\theta)v_i\bigr)
\quad\text{or}\quad
\bigl(i\notin P \text{ and } \tau<(1-\theta)v_i\bigr).
\end{equation}
As before, the algorithm accepts the trigger if a slot remains and
runs Kleinberg's algorithm on subsequent arrivals with the remaining
capacity. Once $P$ and $\tau$ have been computed, this rule no longer
needs the individual predictions.

\begin{theorem}\label{thm:boundary}
With the trigger~\eqref{eq:boundarytrigger} and threshold~\eqref{eq:theta},
the algorithm satisfies
\[
\E[v(A)]\ge
\max\left\{0,1-\min\left\{\frac{32}{\sqrt{k}},4\eta\right\}\right\}\OPT.
\]
If $\eta\le\theta$, no switch occurs and
$v(A)\ge(1-\eta)\OPT/(1+\eta)$. In particular, if $\eta=0$,
the algorithm obtains $\OPT$ on every arrival order.
\end{theorem}

\begin{proof}
Fix $0\le\theta<1$, let $M$ be the set of candidates satisfying
\eqref{eq:boundarytrigger}, and write $m=|M|$ and
$\alpha=(1-\theta)/(1+\theta)$. The key observation is that any
nontriggering shortlisted candidate $i\in P\setminus M$ and any
nontriggering excluded candidate $j\notin P\cup M$ satisfy
\begin{equation}\label{eq:pairwiseboundary}
v_i\ge\frac{\tau}{1+\theta}\ge\alpha v_j.
\end{equation}
This pairwise value comparison replaces the bounds
$(1-\theta)v_i\le\widehat v_i\le(1+\theta)v_i$
used in the proof of Lemma~\ref{lem:Bvalue}.

To see this, suppose $m\le\sqrt{k}$. Let $G=P\setminus M$,
$H=O\setminus M$, and let $H'$ consist of the
$\min\{|G|,|H|\}$ highest-value candidates in $H$.
Since $|G|\ge k-m$ and $|H|\le k$, we have
$v(H')\ge(1-m/k)v(H)$. Match each member of $H'\setminus G$
to a distinct member of $G\setminus H'$; this is possible because
$|H'|\le|G|$. Each matched member of $G$ has value at least
$\alpha$ times that of its partner by~\eqref{eq:pairwiseboundary}.
Members of $G\cap H'$ contribute their own value. Hence
\[
v(G)\ge\alpha v(H')\ge\alpha(1-m/k)v(H).
\]
Adding $v(O\cap M)$ recovers the bound of Lemma~\ref{lem:Bvalue}
for $B=G\cup(O\cap M)$.

The remaining analysis uses only that $M$ is fixed before arrivals
and that the algorithm follows $P$ until the first member of $M$
arrives or capacity is exhausted. Thus the conditional completion,
capacity averaging, and many-error bounds apply unchanged. For
$m=0$, the comparison above gives $v(P)\ge\alpha\OPT$ directly.
This proves Theorem~\ref{thm:robust}
for the modified rule, regardless of $\eta$.

Finally, if $\eta\le\theta$, the strict trigger never fires.
Applying the same matching argument at threshold $\eta$ with
$M=\varnothing$ gives $v(A)=v(P)\ge(1-\eta)\OPT/(1+\eta)$.
Combining this accuracy bound with Theorem~\ref{thm:robust},
which also holds for the modified rule, and following the argument
in the proof of Theorem~\ref{thm:main} yields the stated guarantee.
\end{proof}

\section{Related work}\label{sec:related}

\paragraph{Classical secretary selection.}
The single-choice secretary problem dates to Dynkin~\cite{dynkin1963}; Gupta and Singla~\cite{gupta2021} survey the broader random-order setting. Independent uniform arrival times provide a standard analytical representation~\cite{bruss1984}. For the additive multiple-choice objective, Kleinberg~\cite{kleinberg2005} established the optimal robustness order $1-O(k^{-1/2})$. His algorithm supplies our fallback after a prediction error is detected. Babaioff et al.~\cite{babaioff2007} study knapsack secretary selection, including a $1/e$ value guarantee for equal costs, while Albers and Ladewig~\cite{albers2021} develop simple algorithms with stronger guarantees at small capacities. Qiao and Zhang~\cite{qiao2026} achieve the optimal asymptotic ratio using $O(\log k)$ memory. Their focus is prediction-free memory efficiency; our focus is retaining that robustness order while obtaining the optimum with exact predictions. 

\paragraph{Predictions and advice interfaces.}
Learning-augmented algorithms seek guarantees that combine accuracy-dependent performance with worst-case robustness; see Mitzenmacher and Vassilvitskii~\cite{mitzenmacher2022}. Fujii and Yoshida~\cite{fujii2024} introduce the per-candidate prediction model studied here. Their Learned Kleinberg algorithm achieves loss $O(\min\{\log k/\sqrt{k},\eps\})$; our analysis removes the logarithmic factor with a different monitoring threshold. Antoniadis et al.~\cite{antoniadis2020} predict the optimum value in secretary selection and also study online matching and graphic matroids. Braun and Sarkar~\cite{braun2024} use a predicted additive gap, obtaining a $0.4$ guarantee when that gap is exact. D\"utting et al.~\cite{duetting2024} develop an advice framework and factor-revealing linear programs for success probabilities, encompassing samples, known distributions, and noisy binary advice. Nourmohammadi et al.~\cite{nourmohammadi2026} study \emph{Ordinal Secretaries with Advice}, including multiple-choice and rehiring variants. Their advice concerns the best candidate's arrival position, and their multiple-choice objective is the probability of including that candidate among those selected. Our advice instead gives numerical predictions for labeled candidates, and our objective compares total selected value with the true top-$k$ sum.

\paragraph{Single-choice guarantees and arrival control.}
Karisani et al.~\cite{karisani2026} improve single-choice random-order robustness from $0.215$ to $0.221$, while retaining the accuracy term $(1-\eps)/(1+\eps)$. They obtain $0.262$ robustness when predictions can be used to choose the arrival order. Yiming~\cite{yiming2026} studies prediction-based ordering, with $0.229$ robustness and improved fairness bounds in that setting. Choo and Ling~\cite{choo2024} establish a limiting $1/3$ upper bound for perfectly consistent single-choice algorithms. These works study single-choice robustness; we study how robustness
improves with $k$ under random arrivals.

\paragraph{Fairness and boundary-sensitive error.}
Balkanski, Ma, and Maggiori~\cite{balkanski2024} combine utility guarantees with individual selection-probability guarantees. Their multiple-choice result gives expected value at least $\OPT-4k\Delta$, where $\Delta=\max_i|v_i-\vhat_i|$, together with rank-dependent fairness bounds. We analyze relative total value under multiplicative prediction error. Fujii and Yoshida~\cite{fujii2024} also introduce a boundary-sensitive error: candidates outside the predicted top-$k$ set are compared with its predicted cutoff, while candidates inside retain two-sided multiplicative error. 

\section*{AI Disclosure}
We used ChatGPT to help audit and revise the manuscript.

\bibliographystyle{splncs04}
\bibliography{references}

\end{document}